\documentstyle[aps,graphicx,amsmath,amssymb,bm]{revtex}

\newcommand{\kfn}{k_{F_n}}

\newcommand{\beq}{\begin{equation}}
\newcommand{\eeq}{\end{equation}}
\newcommand{\beqy}{\begin{eqnarray}}
\newcommand{\eeqy}{\end{eqnarray}} 

\newcommand{\si}{{\bm \sigma_1}}
\newcommand{\sit}{{\bm \sigma_2}}

\begin{document}
\draft
\title{Neutrino emission in neutron matter from magnetic moment interactions}

\author{Prashanth Jaikumar, K. R. S. Balaji and Charles Gale}
\address{Physics Department, McGill University, 3600 University Street,
Montr\'eal, Canada H3A 2T8}
\date{\today}
\maketitle

\begin{abstract}
\noindent
Neutrino emission drives neutron star cooling for the first several
hundreds of years after its birth. Given the low energy ($\sim$ keV)
nature of this process, one expects very few nonstandard particle
physics contributions which could affect this rate. Requiring that
any new physics contributions involve light degrees of freedom, one of the
likely candidates which can affect the cooling
process would be a nonzero magnetic moment for the neutrino. To
illustrate, we compute the emission rate for neutrino pair
bremsstrahlung in neutron-neutron scattering through photon-neutrino
magnetic moment coupling. We also present analogous differential rates for
neutrino scattering off nucleons and electrons that determine neutrino
opacities in supernovae. Employing current upper bounds from collider
experiments on the tau magnetic moment, we find that the neutrino
emission rate can exceed the rate through neutral current electroweak
interaction by a factor two, signalling the importance of new particle
physics input to a standard calculation of relevance to neutron star
cooling. However, astrophysical bounds on the neutrino magnetic moment imply
 smaller effects.
\end{abstract}

\section{Introduction}

Neutrino physics plays a crucial role in the birth and subsequent
evolution of neutron stars, beginning with the hot and dense
environment of a supernova~\cite{burrows}, where diffusing neutrinos
are believed to trigger the explosive event, to the interior of a
cold neutron star, whose cooling rate is determined principally by
free-streaming neutrinos~\cite{prakash}.  While various problems
remain with the neutrino-driven supernova explosion mechanism despite
recent intensive efforts (see~\cite{buras} and refs. therein),
uncertainties in neutrino emission from neutron stars is limited to a
lack of knowledge of the precise underlying equation of state of
supra-nuclear matter, where progress is tied to improving many-body
calculations~\cite{schwenk}. The important point is that the long-term
cooling of neutron stars is controlled by neutrino emission, and this
stage lasts up to about $10^5$ years of age, when cooling by emission
of photons becomes more effective.

\vskip 0.2cm

There is a host of well-known neutrino emission processes that operate
in the crust and core of the neutron star. A comprehensive list of
neutrino emitting reactions relevant to different regions of the star
can be found in~\cite{Yakovlev}. Which one dominates the cooling
depends mainly on the temperature and to a lesser extent, on the
density. It should be noted that neutron or proton superfluidity can
also reduce or enhance neutrino emission dramatically through
density-dependent gaps~\cite{gnedin}. Since cooling rates are
dependent on neutrino emissivities, it has proven useful to turn to
neutron star cooling to help identify or constrain new physics
contributions to these emissivities~\cite{Yak03}. A typical example is
the upper bound on the axion mass (or coupling) from axion
bremsstrahlung emission in neutron-neutron collisions~\cite{umeda} and stellar cooling~\cite{georg}.

\vskip 0.2cm

In this work, we are interested in a particular neutrino emission process
involving the neutrino magnetic dipole moment. A magnetic moment for the neutrino was postulated almost immediately following Pauli's neutrino hypothesis. The astrophysical consequences of a neutrino magnetic moment, particularly on stellar cooling have been investigated previously (see~\cite{Iwamoto95} and refs. 3-9 therein).
These works have focused their attention on the plasmon decay process, which
is the dominant cooling mechanism in red giants, and also for the crust of a
neutron star until thermal conduction lowers temperatures significantly to the point when processes occurring in the core become more efficient. Since the crust is only a small volume fraction of the star, specially for softer equations of state, it is important to address neutrino emission from the core which could also receive important contributions from new physics. As an illustrative example of this, we present a calculation of the neutrino emissivity from neutrino bremsstrahlung in neutron-neutron collisions, where a neutron radiates an off-shell photon (via magnetic moment coupling of the neutron), which subsequently decays to a neutrino-antineutrino pair via magnetic moment coupling. Section IIA outlines the general features and expected significance of particle-physics corrections to known neutrino rates. In section IIB, we explain why neutrino bremsstrahlung from neutron-neutron collisions can be the dominant cooling process in the core, and present the computation of the total emissivity from magnetic moment interactions, which can be appreciable. We briefly discuss corrections to scattering rates of neutrinos by electrons and nucleons which dominantly determine neutrino opacities and mean free paths in supernovae environments. The conditions under which new particle physics corrections such as presented here can be quantitatively large are explored in section III. We conclude with a discussion of the relevance of our results in section IV.   

\vskip 0.2cm

\section{neutrino bremsstrahlung through magnetic moment coupling}
\subsection{Motivations for neutrino magnetic moment}
Neutrino pair production could have new nonstandard channels besides the well
known standard model neutral current interactions. We note that for the
present analysis, the typical scale of the process involves momentum
transfers $\sim$ keV which is also the dominant constraint. Therefore, any 
new interactions must involve particles with lighter degrees of freedom.
In this context, one of the more appealing choices of new physics 
is to consider a neutrino magnetic moment interaction with the photon field as the mediator. 

\vskip 0.2cm

 Within the standard model, neutrinos are massless and chiral and
  hence have zero magnetic moment. Neutrinos were first introduced by
  Pauli to explain the energy-momentum conservation in the beta decay
  process, and can also be assigned a magnetic moment.  Carlson and
  Oppenheimer \cite{carlson} and later Bethe \cite{bethe} and
  Domogatskii et. al \cite{domo}, considered the neutrino magnetic
  moment contribution to electron-neutrino scattering.  Existence
  of a non zero magnetic moment can indicate possible masses for the
  neutrino and hence physics beyond the standard
  model~\cite{shrock,marciano}.  If the neutrino interaction was
  purely of the electroweak type, then the neutrino magnetic moment is
  proportional to its mass and is expected to be very small. In other
  words, the chiral nature of neutrinos in such a case will impose a
  small magnetic moment. However, the standard model (which has the
  electroweak sector as a part of it) is considered to be an effective
  low energy theory valid at the electroweak breaking scale of $100$
  GeV.  At larger energy scales, new interactions must be introduced
  and can also lead to values of the magnetic moment larger than the
  standard model expectations. Results from SuperKamiokande on
  neutrino anomalies have generated immense interest in the possibility
  of massive neutrinos, implying physics beyond the standard model
  \cite{sk}.

Analogous to a charged fermion, a neutrino can have an 
anomalous photon-neutrino coupling. This can be defined through an 
electromagnetic current with coupling strength $\kappa$ and momentum transfer 
four vector $q=p^{\prime}-p$. The current is given by
\begin{equation}
J_\mu=i\kappa(q^2)\bar
u(p^{\prime})\sigma_{\mu\nu}q^\nu u(p)~.
\end{equation}

 Here, $\kappa(q^2)$ is the relevant electromagnetic form factor. The energy 
dependence can be extracted by taking the coupling to be of dipole form 
(valid up to a scale $\Lambda$) with the form
  \begin{equation}
  \kappa(q^2) = \frac {\displaystyle \kappa(0)}{\displaystyle
  (1+q^2/\Lambda^2)^2}~.
  \end{equation} 

\vskip 0.2cm

In the present analysis, we shall assume negligible energy dependence;
thus, our constraints can be reinterpreted as those obtained for
$\kappa(0)$. This approximation is valid so long as the physics that
drives $J_\mu$ is at a scale below the TeV region. Then the neutrino
magnetic moment is the value of the form factor at zero momentum
transfer.

\vskip 0.2cm

Nontrivial electromagnetic effects can be expected in light of massive
  neutrinos. In this context, a nonzero transition magnetic moment has
  been used to explain the solar neutrino anomaly observed in
  terrestrial experiments \cite{lim,barbieri,akh}. Furthermore, a
  nonzero $\kappa$ has been used to explain the anticorrelation of the
  solar neutrino flux with the sun spot cycles and the biannual
  variation of the solar neutrino flux \cite{palash}. In the present
  case, we are interested to know if neutrinos with $\kappa \neq 0$
  could as well affect neutron star cooling processes through modified neutrino
  emissivities. Among the many other conceivable sources for new
  physics, we find the neutrino magnetic moment to be one of the more
  attractive examples. The following analysis is to be taken in the spirit of
  illustrating new particle physics contributions to the existing
  standard nuclear physics calculations. There could be many
  other potentially viable options but they are beyond the scope of the
  present work.

\subsection{The neutrino bremsstrahlung process}

For the neutrino emissivity, we will focus on neutrino bremsstrahlung
from neutron-neutron collisions alone (pure neutron matter), although
similar effects can easily be computed in neutron-proton scattering as
well. The reason is that the latter, as well as the standard cooling
process, namely the modified Urca ($n+n\rightarrow~n+p+e^-+\bar{\nu}_e$), is
strongly suppressed due to proton superfluidity, whereas, at the
relevant core temperatures, the neutrons are likely to be in the
normal phase. This conclusion follows , given that $^1$S$_0$
superfluidity of neutrons ceases to exist (due to $s-$wave repulsion
above nuclear matter density) and that triplet gaps are much weakened
due to strong renormalization of non-central interactions in-medium at
or above nuclear matter density~\cite{schwenk03}. In this case, the
dominant neutrino emission process will be neutrino bremsstrahlung in
$nn$ collisions.  In fact, even at lower densities, where both
neutrons and protons pair in the $^1$S$_0$ channel, the neutrino pair
emissivity can be more effective compared to the modified Urca
process~\cite{Yakovlev}. With this motivation, we proceed to estimate
the neutrino emissivity from $nn$ bremsstrahlung.

\vskip 0.2cm

The Feynman diagrams that contribute at tree level to the neutrino
pair bremsstrahlung in $nn$-scattering involve photon emission from each of the four neutron legs, with the photon subsequently coupling to a neutrino pair via the magnetic moment (see eqn.(\ref{magint}) below). For the nuclear matrix
element, we use the results of Friman and
Maxwell~\cite{FM} which take into account the long-range one-pion exchange
tensor force in the nucleon-nucleon scattering amplitude explicitly
and estimate the effects of short-range correlations by cutting off
the interaction at short distances and including the short-range
rho-exchange tensor force.~\footnote{We have not included the
exchange terms and the repulsive short-range rho-exchange, since, as
shown in~\cite{FM}, these effects largely offset each other. We have also ignored photon emission from the exhanged pion, because, unlike the intermediate neutron leg, the former does not lead to a small energy denominator. (see eqn.(\ref{smallomega}))} More
recently, revised bremsstrahlung rates in neutron matter that
incorporate many-body effects in the medium seem to point to a strong
reduction of the emissivity at sub-nuclear and nuclear densities
compared to those results~\cite{jaikumar}. As we are interested only in the
relative importance of new particle physics input to the emissivity,
we will, for the sake of simplicity, compare our result to the
benchmark calculation in~\cite{FM}. 
\vskip 0.2cm

In the notation of~\cite{FM}, the emissivity is given by (for $\hbar=c=1$)

\beq
\label{emissmain}
\varepsilon_{\gamma\nu\bar{\nu}} = N_{\nu} \int \biggl( \, \prod_{i=1}^4 \frac{d^3
{\bm p}_i}{(2\pi)^3} \biggr) \, \frac{d^3 {\bm
Q}_1}{2\omega_1(2\pi)^3} \, \frac{d^3 {\bm Q}_2}{2\omega_2(2\pi)^3} \,
(2 \pi)^4 \, \delta(E_f-E_i) \, \delta^3({\bm P}_f - {\bm P}_i) \,
\frac{1}{s} \, \biggl( \: \sum_{\text{spin}} \, \bigl| \,
{\mathcal{M}}_{nn} \bigr|^2 \biggr) \, \omega_{\nu} \,
{\mathcal{F}}(E_{{\bm p}_i})\,, 
\eeq 

where ${\bm p}_i$ denote the
momenta of the incoming and outgoing neutrons and
$Q_{1,2}=(\omega_{1,2},{\bm Q}_{1,2})$ label the neutrino energies and
momenta. The delta functions account for energy and momentum
conservation, and $\omega_{\nu}=\omega_1+\omega_2$ is the total
neutrino energy.  $N_{\nu}$ denotes the number of neutrino species and
$s=2$ is a symmetry factor for the initial neutrons, when the emission
occurs in the final state or vice versa.  The function
${\cal{F}}(E_{{\bm p}_i})=f(E_{{\bm p}_1}) \, f(E_{{\bm p}_2}) \,
(1-f(E_{{\bm p}_3})) \, (1-f(E_{{\bm p}_4}))$ is the product of
Fermi-Dirac distribution functions $f(E)=(\exp(E/T) + 1)^{-1}$, with
neutron energies $E_{{\bm p}_i}$.  The matrix element
${\mathcal{M}}_{nn}$ includes nucleon-nucleon scattering and
the coupling to the emitted neutrino pair. The former is described by
one-pion exchange, 

\beq V_{\rm
OPE}=\biggl(\frac{f}{m_{\pi}}\biggr)^2\si\cdot{\bf
k}\biggl(\frac{-1}{{\bf k}^2+m_{\pi}^2}\biggr)\sit\cdot{\bf
k}({\bm \tau}_1\cdot{\bm \tau}_2) \,,
\eeq 

where the pion-nucleon coupling constant
$f^2\approx 4\pi\times 0.08\approx 1$, and ${\bm \sigma}$ denotes spin
and ${\bm \tau}$ isospin respectively.  ${\bf k}$ is the momentum exchanged in the scattering event, equal to ${\bm p}_1-{\bm p}_3$ (or ${\bm p}_4-{\bm p}_2$) . The emitted photon couples to
the neutron through the magnetic dipole moment, (denoted by
$\kappa_n$). Since the neutrons are non-relativistic, and the emitted
neutrinos are thermal, to ${\cal O}({\bm Q}/m_N)$ (where ${\bm Q}$ is
the three-momentum of the neutrino pair), we may write the
neutron-photon coupling as 

\beq 
i\kappa_n\xi^{\prime}\frac{{\bm
\sigma}\times{\bm Q}}{m_N}\xi \,,
\eeq 

where $\xi,\xi^{\prime}$ are incoming and outgoing non-relativistic
free nucleon spinors normalized to unity. All other non-relativistic
reductions of the fully relativistic neutron-photon vertex couplings
are negligible (of higher order in ${\bm Q}/{m_N}$ in the limit of small $Q^2$). In this limit,
$\kappa_n=-1.91$ is simply the ratio of the magnetic Sachs factor to
the Pauli moment of the neutron. The photon current couples to a
neutrino pair also via the neutrino magnetic moment (denoted by
$\kappa$): 

\beq
\langle\nu(Q_2)|J^{\mu}_{\gamma}|\nu(Q_1)\rangle=i\kappa\bar{u}(Q_2)\sigma^{\mu\nu}(Q_2-Q_1)_{\nu}{u}(Q_1)\,,\label{magint}
\eeq 

where $u$'s are normalized relativistic spinors for massless neutrinos, and
$\sigma^{\mu\nu}=\frac{i}{2}[\gamma^{\mu}\gamma^{\nu}]$. In computing the matrix element, we have, as
in~\cite{FM}, used a non-relativistic approximation for all nucleon
propagators, where the lowest term in an expansion in inverse powers
of the nucleon mass is retained. In addition, one neglects the
neutrino pair energy $\omega_\nu$ compared to the Fermi energy, since
the emitted neutrinos are thermal and have negligible energy compared to the neutrons at typical neutron star densities and temperatures. For the nucleon propagator $G$,
this approximation yields 

\beq i \, G({\bm p} \pm {\bm Q},E_{\bm
p} \pm \omega_\nu) = \pm i \, \omega_\nu^{-1} ,
\label{smallomega}
\eeq 

where the positive sign holds if the electromagnetic current is attached to
an outgoing nucleon, negative otherwise. The spin trace over the 
neutrino pieces yields

\beq
{\rm Tr}(l_{\mu}l_{\nu})=-2g_{\mu\nu}s^2+8g_{\mu\nu}(Q_1.Q)(Q_2.Q)+s[4Q_{1\mu}Q_{2\nu}+4Q_{1\nu}Q_{2\mu}+2Q_{\mu}Q_{\nu}]-4[(Q_1.Q)(Q_{2\mu}Q_{\nu}+Q_{2\nu}Q_{\mu})+Q_1\leftrightarrow Q_2]
\eeq

where $s=Q^2=(Q_1+Q_2)^2$ is a kinematic variable (center-of-mass momentum squared). Terms anti-symmetric in $\mu$ and $\nu$ are dropped since they eventually contract to zero. 
The non-relativistic reduction of the neutron-photon vertex implies that only the spatial indices (such as $\mu=i,\nu=j$) contribute. Furthermore, by transversality of the photon, the pieces of the neutrino spin summed matrix element that are proportional to $Q_i$ vanish on contraction with the nuclear spin summed matrix element squared, leaving us with

\beqy
\langle\bigl|{\mathcal{M}}_{nn}\bigr|^2\rangle&=&128\frac{\kappa^2\kappa_n^2}{m_N^2\omega^2}\biggl(\frac{f}{m_{\pi}}\biggr)^2\biggl(\frac{{\bf k}^2}{{\bf k}^2+m_{\pi}^2}\biggr)^2\frac{{\bm Q}^2}{s}\\ \nonumber 
&&\times[2({\bm Q}_1\cdot{\bm Q}_2)({\hat {\bm Q}}\cdot{\hat {\bm k}})^2+2({\bm Q}_1\cdot{\hat {\bm k}})({\bm Q}_2\cdot{\hat {\bm k}})-2({\hat {\bm Q}}\cdot{\hat {\bm k}})\{({\bm Q}_1\cdot{\hat {\bm k}})({\bm Q}_2\cdot{\hat {\bm Q}})+({\bm Q}_1\cdot{\hat {\bm Q}})({\bm Q}_2\cdot{\hat {\bm k}})\}]\,,
\eeqy
  
where an averaging over initial neutron spins has been performed. Note that there is a $1/s$-dependence in the above expression, which indicates an enhancement at small $s$. There is no divergence, however, since the quantity in the bracket [..] vanishes when the photon is on-shell ($s=0$); the decay is then kinematically forbidden. Furthermore, when the integrations over the neutrino momenta are performed, an extra factor of $s$ appears in the numerator to cancel the one in the denominator. We may note here that a $1/t$ enhancement does appear in the differential cross-section for the $t-$channel process of electron-neutrino scattering which is related to a forward peak in the differential cross-section~\cite{Barut}. For the differential emissivity, which is a function of the energies of the outgoing neutrinos and the relative angle between them, we have checked that there is a forward peak in the $s-$channel as well, although the $1/s$ enhancement is canceled by the angular phase space measure. Differential neutrino rates are known to be important in the microphysics of neutrino transport in a supernova~\cite{rat}, in which case the forward enhancement in the scattering channel can contribute significantly to thermalization and neutrino opacities. We comment on these toward the end of this section. Proceeding with the calculation for the total emissivity, the neutrino phase space integrals are easily performed by introducing delta functions 
as 
\beq
1 = \int d\omega_{\nu}\,\delta(\omega_{\nu}-\omega_1-\omega_2)\,,  1 = \int d^3{\bm Q} \delta({\bm Q}-{\bm Q}_1-{\bm Q}_2) \,,
\eeq

and then using Lenard's identity
\beq
N_{\alpha\beta}=\int
\hspace{0.05in} \frac{d^{3}Q_{1}}{2\omega_1} \frac{d^{3}Q_{2}}{2\omega_2}~
(Q_{1\alpha}Q_{2\beta}+Q_{2\alpha}Q_{1\beta}) 
\hspace{0.05in} \delta^{4}(Q_{1} + Q_{2}-Q)=\frac{\pi
}{12}(Q^{2}g_{\alpha\beta} +
2Q_{\alpha}Q_{\beta})\Theta(Q_{0})\Theta(Q_{0}^{2} -
{\bm Q}^2) \,,
\eeq 
Next, one can decouple the angular parts in the
neutron phase space and trade the radial momentum for energy integrals,
by restricting the interacting neutrons to the surface of the Fermi sphere (of radius $\kfn$), since they 
are strongly degenerate for typical neutron star temperatures. Corrections
to this approximation scale as $T/E_{F_n}$ where $E_{F_n}$ is the Fermi temperature (in $k_B=1$ units). For this
purpose, one replaces
\beq 
d^3 {\bm p}_i \rightarrow d^3 {\bm p}_i \, \frac{m_n^\ast}{\kfn}
\, \delta(p_i - \kfn) \, \int dE_{{\bm p}_i}\,.
\eeq 
where $m_n^\ast$ is the neutron effective mass. After carrying out the angular integrations, we find a compact expression 
for the emissivity
\beqy
\varepsilon_{\gamma\nu\bar{\nu}}&=&\frac{A}{(2\pi)^{12}}I_{\nu\bar{\nu}}\biggl(\frac{m_n^\ast}{\kfn}\biggr)^4{\cal S}\,,\\
A&=&\frac{32}{45}\biggl(\frac{f}{m_{\pi}}\biggr)^4\biggl(\frac{\kappa\kappa_n}{m_n}\biggr)^2\,, \,I_{\nu\bar{\nu}}=\frac{41}{60480}(2\pi)^8T^8 \,, \\
S&=&\int\biggl[\prod_{i=1}^{4}d^3{\bm p}_i\delta(p_i - \kfn)\biggr]\biggl(\frac{{\bf k}^2}{{\bf k}^2+m_{\pi}^2}\biggr)^2\delta^3({\bm P}_f-{\bm P}_i)\,.
\eeqy
${\cal S}$ is the neutron phase space integral which is facilitated by introducing the integration over momentum transfers through
respective delta functions as
\beq
1 = \int d^3{\bm k} \, \delta^3({\bm k} - {\bm p}_1 + {\bm p}_3)\,.
\eeq
The neutron phase space integrals yield~\cite{FM}
\beqy
{\cal S}&=&4(2\pi)^3\kfn^5\biggl(\frac{f}{m_{\pi}}\biggr)^4F\biggl(\frac{m_{\pi}}{2\kfn}\biggr)\,,\\
F(x)&=&1-\frac{3}{2}{\rm tan}^{-1}\biggl(\frac{1}{x}\biggr)+\frac{1}{2}\biggl(\frac{x^2}{1+x^2}\biggr)\,.
\eeqy
It is convenient for purposes of comparison to express the result for the emissivity as a ratio to that obtained from bremsstrahlung via electroweak coupling to the neutrino, whose value $\varepsilon_{\nu\bar{\nu}}$ is calculated in~\cite{FM}:
\beq
\frac{\varepsilon_{\gamma\nu\bar{\nu}}}{\varepsilon_{\nu\bar{\nu}}}=\frac{4}{3}\biggl(\frac{\kappa\kappa_n}{G_Fg_Am_n^\ast}\biggr)^2\,.
\eeq
Numerically, we find this ratio to be ($G_F=1.166\times 10^{-5}$~GeV$^{-2}, g_A=1.26, \kappa_n=-1.91$)  
\beqy
R^{nn}=\frac{\varepsilon_{\gamma\nu\bar{\nu}}}{\varepsilon_{\nu\bar{\nu}}}=2.25\biggl(\frac{\kappa}{10^{-5}{\rm GeV}^{-1}}\frac{1}{m_n^\ast}\biggr)^2\,,\label{rnn} \eeqy 
where $m_n^\ast$ is taken in units of GeV.\\

A similar computation for the $np\rightarrow np\nu\bar{\nu}$ bremsstrahlung channel which is more important in the stellar crust yields ($\kappa_p=1.79$)
\beq
R^{np}=\frac{2}{3}\biggl(\frac{\kappa}{G_Fg_A}\biggr)^2\biggl[\frac{\kappa_n^2}
{m_n^{\ast^2}}+\frac{(1+\kappa_p)^2}{m_p^{\ast^2}}\biggr]=0.31\biggl(\frac{\kappa}
{10^{-5}{\rm GeV}^{-1}}\biggr)^2\biggl[\frac{(-1.91)^2}{m_n^{\ast^2}}+\frac{(2.79)^2}
{m_p^{\ast^2}}\biggr]\,. 
\label{ratio}
\eeq
The comparative equations (\ref{rnn}) and (\ref{ratio}) are also valid for bremsstrahlung from non-degenerate nucleonic matter (the weak interaction rates for which are given in~\cite{Thomson00}), in the approximation that Pauli blocking of neutrinos can be neglected. Thus, the above estimates hold as one progresses outwards from the neutrinosphere of a young protoneutron star or a supernova as well. In the cores of these objects, neutrinos have short mean free path, and their opacity and thermalization time scale is controlled by (for $\nu_{\mu}$ and $\nu_{\tau}$ neutrinos) processes such as $\nu-$nucleon and $\nu-$electron scattering. If a neutrino magnetic moment exists, given the $1/t$ enhancement in the scattering channel, one can expect magnetic moment interactions to dominate over weak interactions which do not have this feature in the differential cross-section. This would impact on neutrino transport and equilibration, which is a key component of theoretical models of supernovae explosion. In this work, we have chosen to focus on the emission process (bremsstrahlung) to highlight the possible importance of non-standard contributions to neutrino astrophysics, but we can also outline the effects on scattering processes.

\vskip 0.2cm

In the cores of protoneutron stars and supernovae, typical ambient temperatures ($T\sim 1-50$ MeV) and densities ($\varrho\sim 10^{10}-10^{14}$ g/cc) imply that neutrinos have short mean free paths due to frequent scattering off electrons and nucleons, which are the dominant sources of neutrino opacity. In both cases, differential rates have been computed, which serve as the particle-physics input to the Boltzmann equation that in turn evolves the neutrino phase-space distribution (${\cal F_{\nu}}$). For example, assuming a homogeneous, isotropic thermal bath of scatterers and absorbers, the governing equation for neutrino transport is
\beq
\frac{\partial {\cal F_{\nu}}}{\partial t}=(1-{\cal F_{\nu}})J_{\nu}-{\cal F_{\nu}}A_{\nu}\,.
\eeq
where $J_{\nu}$ and $A_{\nu}$ are respectively the neutrino emission and absorption currents from one energy bin to another. These quantities are dependent on the differential cross-section for inelastic neutrino scattering. However, since we are only interested in the relative importance of the magnetic moment effects, we may compare elastic differential cross-sections for simplicity. This is equivalent to assuming a redistribution of neutrinos only in space and not in energy. For a rigorous treatment of inelastic rates and their incorporation into the Boltzmann equation, one can follow the structure function formalism developed in ~\cite{RPL}. Below, we will only compare elastic rates as they will suffice to estimate the magnitude of the corrections through magnetic moment interactions. 

\vskip 0.2cm

For neutrino-nucleon ($N=n$ or $p$) scattering, the differential rates for elastic scattering as a function of the angle between the incident and scattered neutrino $\theta$ and the neutrino energy $E_{\nu}$ are given by 
\beqy
\frac{d\sigma^{\nu n\rightarrow\nu n}}{d{\rm cos}\theta}&=&\frac{G_F^2E_{\nu}^2}{2\pi}\biggl((1+{\rm cos}\theta)+g_A^2(3-{\rm cos}\theta)\biggr)\,,\\ \nonumber 
\frac{d\sigma^{\nu p\rightarrow\nu p}}{d{\rm cos}\theta}&=&\frac{G_F^2E_{\nu}^2}{8\pi}\biggl((16\alpha^2-8\alpha+1)(1+{\rm cos}\theta)+g_A^2(3-{\rm cos}\theta)\biggr);\quad \alpha={\rm sin}^2\theta_W=0.23\,.
\eeqy
which can be compared against the result from the magnetic moment interaction
\beqy
\frac{d\sigma^{\nu N\rightarrow\nu N}}{d{\rm cos}\theta}&=&\frac{\beta_N^2\kappa^2E_{\nu}^4}{2\pi~E_N^2}{\rm sin}^2\biggl(\frac{\theta}{2}\biggr)\,,\\ \nonumber
\beta_N&=&\frac{\kappa_n}{m_n^{\ast}} {\rm ~for~ neutrons},\quad \frac{1+\kappa_p}{m_p^{\ast}} {\rm ~for ~protons}\,.
\eeqy
Assuming thermal neutrinos with an energy $E_{\nu}= T$ (in $k_B=1$ units), a nucleon effective mass of 900 MeV and $\kappa=10^{-7}~\mu_B$, the ratio (denoted $R_{n/p}$ respectively) of the magnetic moment induced scattering rate to the electroweak rates is displayed as a function of the scattering angle in the figure below. We employ two sets of typical temperatures, densities, nucleon and electron fractions obtained in a one-dimensional core-collapse model, Star~\cite{Thomson00}. We have scaled the proton ratio $R_p$ at the higher temperature of $T=10.56$ MeV by a factor 1/10 to fit within the graph. It is evident that the magnetic moment corrections are more important in the case of neutrino scattering off protons, particularly at  temperatures of 5 MeV or more. 

\vskip 0.2cm

For neutrino-electron scattering, the elastic differential cross-sections ignoring polarization effects of the medium are given by
\beq
\frac{d\sigma_{EW}^{\nu e^-\rightarrow\nu e^-}}{d{\rm cos}\theta}=\frac{G_F^2}{2\pi~s}\frac{(s-m_e^2)^2}{(1-t/M_Z^2)^2}\,,\quad \frac{d\sigma_{MAG}^{\nu e^-\rightarrow\nu e^-}}{d{\rm cos}\theta}=\frac{\alpha_e\kappa^2}{2}\frac{(s-m_e^2)(m_e^2-s-t)}{st}\,.
\eeq
where $s=(E_{\nu}+E_e)^2, t=-2E_{\nu}^2(1-{\rm cos}\theta)$. A comparison of these rates, denoted by the curve $R_e$, is also presented in the same figure, for $T=4.52$ MeV. There is a sharp forward peak which comes from the enhancement in magnetic moment scattering at small momentum transfer (or small angles). This leads to a logarithmic divergence in the total cross-section $\sigma_{\rm tot}^{MAG}\sim \alpha_{\rm em}\kappa^2{\rm log}(t_{\rm max}/t_{\rm min})$, which can be tamed by choosing typical finite cutoffs, for eg. $t_{\rm max} \sim$ MeV, $t_{\rm min} \sim$ keV.  Although these total cross-sections are small compared to the electroweak cross-section $\sigma_{\rm tot}^{EW}\sim G_F^2s/\pi$, and we have largely ignored the fact that the nucleons/electrons experience strong medium effects, these estimates point to significantly increased differential rates for forward scattering (and possible energy redistribution from inelastic rates) upon inclusion of magnetic moment effects for neutrino diffusion in supernovae. 
 
\begin{figure}[t]
\begin{center}
\includegraphics[scale=0.5,clip=,angle=270]{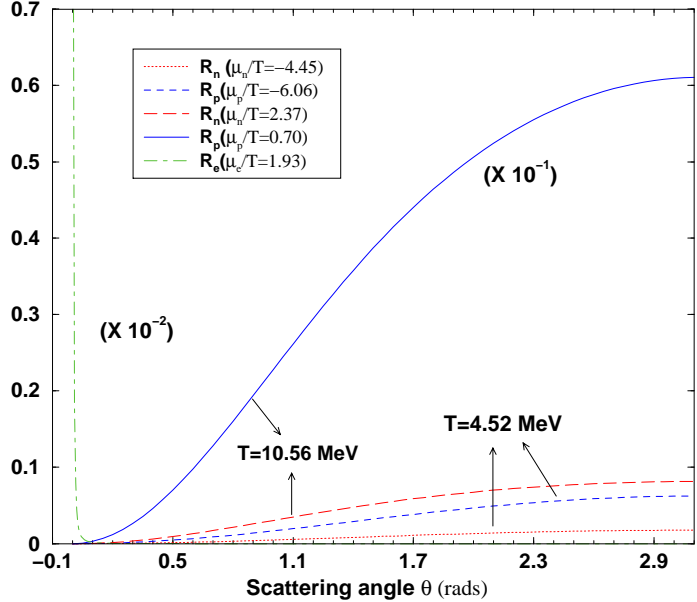}
\end{center}
\caption{Ratio of differential rates from magnetic moment interactions ($\kappa=10^{-7}~\mu_B$) to electroweak, for $\nu-n$ and $\nu-p$ scattering. The $\nu-p$ ratio at $T=10.56$ MeV is scaled by 0.1 to fit within the plotted range. The dot-dashed line is the ratio, scaled by a factor .001, in case of $\nu-e^-$ scattering (peaked at $\theta=0$) for $T=4.52$ MeV. The zero on the x-axis is slightly offset to show the peak.}
\label{diag}
\end{figure}
\vskip 0.2cm

\section{contribution to neutrino bremsstrahlung emissivity due to $\kappa$} 

We can assess the importance of these effects by utilizing the upper bounds on the $e,\mu,$ and $\tau$ neutrino magnetic moments coming from collider experiments and astrophysical data on stellar cooling.    
The process $e^+e^-~\rightarrow ~ \nu \overline{\nu} \gamma $ 
has been used to examine the constraints on the photon induced 
electromagnetic interaction for the neutrino ($\kappa$). Earlier, bounds for $\kappa$ 
were obtained at the Z pole \cite{gould} and also below the $Z$ pole \cite{desh}.
Currently, both  the $\nu_e$ (from SN1987A data) and
  $\nu_{\mu}$ (from $\nu_{\mu}e^-$ scattering experiments) set strong
  bounds on $\kappa$, limiting it to less than
  $10^{-10}\mu_B$. Using the solar neutrino data from
  Kamiokande, a bound on $\kappa~(\sim 10^{-10}\mu_B)$ was
  obtained where an energy independent
  suppression of the $\nu_e$ flux was assumed \cite{suzuki}. Such an
  approach is now ruled
  out considering the fact that the $\nu_e$ flux depletion is indeed
  energy dependent.
Assuming three active neutrino flavors, as
  suggested by the $Z$ width measurements at LEP, only the
  $\nu_{\tau}$ is yet not so strongly constrained with respect to its
  anomalous electromagnetic couplings.  Recently, the 
  atmospheric neutrino data from
  Super-Kamiokande has been used to achieve bounds on $\nu_{\tau}$
  magnetic moment $(\kappa \sim 10^{-7} \mu_B)$ assuming
 maximal $\nu_{\mu}
  \leftrightarrow \nu_{\tau}$ mixing \cite{gninenko} and is three orders
  of magnitude weaker than the collider bounds. However, given that the
mass of the $\nu_\tau$ is not expected to be very different from other neutrino
flavors, one could extrapolate the astrophysical limits (which are flavor
blind) to the case of $\nu_\tau$ as well. In this case, we obtain values
of $\kappa \leq 10^{-10} \mu_B$ \cite{pdg}.  For the present purposes, we 
therefore consider the range $10^{-10} \mu_B \leq \kappa \leq 10^{-7} \mu_B$ and we 
shall show how this variation can affect the relative 
contribution to the neutron star cooling from standard expectations.
To illustrate, we use the following conversions:
$1 \mu_B = 5.8 \cdot 10^{-9}$ eV/Gauss and (ii) 1Gauss = $7 \cdot 10^{-2}$eV$^2$
and therefore, $1\mu_B = 8.2 \cdot 10^{-8}$ eV$^{-1}$~.
Using (\ref{ratio}) and setting the nucleon masses to be of 0.9 GeV, we obtain 
the following limits:

\beq
1.86\cdot 10^{-6}\leq R^{nn}\leq 1.86,\quad 3.61\cdot 10^{-6} 
\leq R^{np} \leq 3.61~\mbox{for the range}~\, 10^{-10} \mu_B \leq \kappa \leq 
 10^{-7} \mu_B ~.
\label{range}
\eeq
Clearly, the upper limit indicates that the new physics can alter the standard 
model-based result for the bremsstrahlung emissivity by a factor of 2-3 while the more conservative lower limit will have negligible effects and we may ignore any particle physics corrections to standard results. The important point is that we cannot rule out sizeable corrections to neutrino bremsstrahlung rates from physics beyond the standard model.  

\section{conclusions}

We have explored the relevance of particle physics beyond the standard model to standard neutrino pair emission rates from neutron star interiors, focusing especially on the role of the neutrino magnetic moment. General arguments based on relevant energy scales imply that a neutrino magnetic moment provides the most interesting and feasible correction to known physics. In the low-energy approximate treatment of neutrino pair bremsstrahlung from neutron matter, we have estimated the correction to a benchmark calculation of the neutrino pair emissivity from electroweak interactions in neutron matter. The magnitude of the correction is estimated from limits on the neutrino magnetic moment $\kappa$. \\

While stringent collider bounds on the $\nu_e$ and $\nu_{\mu}$ electromagnetic moments exist, the $\nu_{\tau}$ is not as nearly strongly constrained, and employing a range set by collider and astrophysical inputs, we find that the standard emission rates may be revised by upto a factor of two or more. We expect similar results for neutrino scattering via magnetic moment interactions, leading to modified neutrino opacities. We have estimated the magnitude of such corrections by using elastic rates and free scatterers. For practical calculations in supernovae physics, we need to explore similar corrections to inelastic rates, which requires including medium effects and energy transfer. These studies are currently in progress. The magnetic moment interaction vertex remaining unaffected in medium, we would expect comparable enhancements upon inclusion of medium effects, therefore we have, in this work, presented ratios of rates rather than their absolute values. We observe that large corrections to the differential cross-section are obtained for neutrino-proton scattering over a wide range of angles and neutrino-electron scattering in the forward direction. The correction to neutrino-neutron scattering is expected to be at the few percent level.

\vskip 0.2cm 

It is also pertinent to mention here the consequences of these revised emissivities for the temperature versus time profile (cooling curve) of a typical neutron star. While stellar cooling is determined (for late times) principally by neutrino emission from the core, a temperature profile requires mapping the surface temperature to the core temperature, a procedure that is strongly dependent on important factors such as local surface temperature variations, magnetospheric emission as well as the composition of the stellar envelope and atmosphere. These details form an integral part of the surface temperature determination. Furthermore, knowledge of the temperature gradient within the core and upto the star's surface requires as input, the equation of state of nuclear matter at supra-nuclear density, and knowledge of related variables such as the incompressibility and the symmetry energy, which are poorly constrained in neutron matter. Model-dependent theoretical uncertainties tend to mask the effects of improved or revised estimates of standard emissivities when incorporated into a numerical simulation of neutron star cooling, and observational difficulties complicate the direct comparison of such numerical results to data. This uncertainty may be removed if a novel temperature or density dependence is discovered as in the case of the direct urca process or if exotic phases such as pion or kaon condensates exist from which cooling can be more efficient. It is unlikely that particle physics corrections to emissivities that are of ${\cal O}(1)$ would considerably alter the overall cooling scenario. Nevertheless, the interesting possibility remains that large contributions from non-standard model physics may exist and yet remain hidden in the microphysics of neutron star cooling.           
\section*{Acknowledgments} 
We thank John Beacom, Walter Grimus and Georg Raffelt for useful
communications. This work was supported by the Natural Sciences and
Engineering Research Council of Canada and the Fonds Nature et
Technologies of Quebec.

\end{document}